\documentclass{pasj00}

\begin{document}
\SetRunningHead{H. Inoue and N. Y. Yamasaki}{X-Ray Detection from SFB in 
Protp-Elliptical Galaxies}
\Received{2002/09/05}
\Accepted{2003/04/23}

\title{An X-Ray Detection Possibility of 
Star-Formation-Bursting Proto-Elliptical Galaxies}

\author{Hajime \textsc{Inoue} and Noriko Y. \textsc{Yamasaki}}
\affil{The Institute of Space and Astronautical Science,
3-1-1,Yoshinodai,Sagamihara,Kanagawa 229-8510}
\email{inoue@astro.isas.ac.jp}


\maketitle

\begin{abstract}
A possibility to detect X-rays from star-formation burst activities 
in proto-elliptical galaxies is considered.  The X-ray flux of an 
emission due to inverse Compton scattering of the cosmic microwave 
background (CMB) by high energy electrons accelerated in SNRs is shown 
to increase as $z$ increases far beyond unity, since the local CMB flux 
largely increases in association with a $z$-increase.   The flux is estimated 
for the case of a very high rate of type II supernovae at an initial star 
formation burst of a proto-elliptical galaxy and is found to be detectable 
with a future large X-ray telescope such as intended in the XEUS mission.  
\end{abstract}

\section{Introduction}

X-ray observations of clusters of galaxies revealed the presence 
of large amounts of iron \citep{mitchel75,serlemitsos76}
and other heavy elements \citep{mushotzky96}
in the intracluster medium (ICM).  
Supernova-driven galactic winds, which were first considered in detail 
by \citet{larson74}, 
are now widely believed to be the origin of ICM enrichment.  

Based on this hypothesis, several constraints on contributions 
of various galaxies to the ICM enrichment and 
on the past supernova (SN) activities in the galaxies 
were obtained by comparing observations with theoretical expectations.
The mass of ICM, which is proportional 
to the iron mass, highly correlates with the luminosity from ellipticals 
and lenticulars and its correlation with that of spirals is poor 
\citep{arnaud92}.  This indicates that only ellipticals and 
lenticulars have participated in the iron enrichment of the ICM.
On the other hand, \citet{renzini93} argued 
that the observed iron mass-to-light ratio cannot be explained unless 
the past average rate of Type Ia supernova was at least a factor of 
$\sim$ 10 higher than the present rate in ellipticals, 
or massive stars in clusters formed 
with a very flat initial mass function.
A similar conclusion was obtained by comparing the total observed amounts 
of not only iron, but also oxygen 
and silicon with calculated yields from type II 
supernovae \citep{loewenstein96}.
\citet{musholoe97}
investigated the $z$-dependence of the iron 
abundance in clusters of galaxies, and found little or no evolution of it 
out to $z \sim$ 0.5.  
By combining this with a passive evolution of elliptical galaxies 
known at least up to $z \sim$ 0.3 -- 0.4, most 
of the iron enrichment should have occurred at $z >$ 1.

\citet{elbaz95} proposed a detailed model for  bimodal star 
formation in elliptical galaxies and the enrichment of the ICM, 
that is, at least, consistent with the above constraints from 
the observations.
In this model, it is considered that about 10$^{10}$ 
type II supernovae occurred during the initial star-formation burst 
for about 10$^{7.5}$ years in proto elliptical galaxies at around 
$z \sim$ 5 -- 10, and that most of iron was produced then.

If this very high rate of SN II explosions really happened in proto 
elliptical galaxies at a very early stage of the universe, 
trials to obtain its observational evidences should be very important.
In this paper, we present an interesting possibility to find them with 
X-ray observations.

Advanced Satellite for Cosmology and Astrophysics 
(ASCA: \cite{tanaka94}) found a firm evidence of presence of very high energy 
electrons in supernova remnants (SNRs) from SN 1006 \citep{koyama95}.  
It was further confirmed by a detection 
of TeV $\gamma$-rays from SN 1006 with the CANGAROO Cerenkov light telescope 
\citep{tanimori98}.
The TeV $\gamma$-rays are well explained as the inverse-Compton scattering 
of the cosmic microwave background (CMB) by the same high energy electrons 
as are responsible for the Synchrotron emission in the radio and X-ray bands.
In relation to the origin of the cosmic-ray particles, such a presence of 
high-energy particles as in SN 1006 would be common to every SNR.  
If so, we would expect an enhancement of the inverse-Compton emission of 
the CMB from an SNR at an early epoch of the universe because of 
the (1$+ z)^{4}$ 
dependence of the CMB flux on the cosmological redshift factor $z$.
The importance of the inverse-Compton emission of the CMB in the early universe 
has been pointed out by \citet{oh01}.
Hereafter, we consider the detection possibility of the inverse-Compton emission 
of the CMB from type II SNRs in star-formation-bursting proto-elliptical galaxies 
in the early universe.

\section{Emission from SNRs Due to Inverse-Compton Scattering of Cosmic Microwave Background}

Let us consider the case that a hot plasma with thermal energy, $U_{\rm Th}$, was 
produced through  shock, and 
that a fraction, $f$, of $U_{\rm Th}$ was transferred to the total energy of 
non-thermal protons, $U_{\rm NT, p}$:  
\begin{equation}
U_{\rm NT, p} = f U_{\rm Th}.
\end{equation}
When the distribution functions in terms of the Lorentz factor, $\gamma$, 
of the non-thermal protons and electrons in $\gamma > 1$ 
are expressed as 
$N_{\rm NT, p} = C_{\rm p}\gamma^{-\mu}$ and 
$N_{\rm NT, e} = C_{\rm e}\gamma^{-\mu}$, respectively, 
the two normalization factors are related to each other as 
$C_{\rm e} = (m_{\rm e}/m_{\rm p})^{(\mu-3)/2} C_{\rm p}$, 
according to \citep{bell78}, where $m_{\rm e}$ and $m_{\rm p}$ are 
the electron and proton mass, respectively.
Thus, the normalization factor of the electron distribution, $C_{\rm e}$, 
can be given with the help of equation (1) as
\begin{equation}
C_{\rm e} =(m_{\rm e}/m_{\rm p})^{-1/2} f U_{\rm Th} /[m_{\rm p} c^{2} 
\ln(\gamma_{\rm p, Max})],
\end{equation} 
for $\mu=2$, since $U_{\rm NT, p} = C_{\rm p}m_{\rm p}c^{2}\ln(\gamma_{\rm p, Max})$.  
Here, $c$ is the velocity of light.

In an SNR, these non-thermal electrons will collide with 
the CMB photons, and radiate inverse-Compton emission.
Since the photon energy, $E$, after a collision of a CMB photon with 
an average energy, $2.7 ~kT_{\rm CMB}$, 
with an electron with $\gamma$ is approximated as
\begin{equation}
E \simeq (4/3)\gamma^{2}(2.7~kT_{\rm CMB}), 
\end{equation} 
electrons with $\gamma \sim 10^{3}$ are responsible for X-ray emission.
Here, $k$ is the Boltzman constant and 
$T_{\rm CMB}$ is the blackbody temperature of the CMB. 
These electrons will lose their energy mainly through ionization loss,  
and their life, $\tau_{\rm e, IL}$ is roughly given as 
(for the ionization loss rate, see e.g., \cite{ginzburg79})
\begin{equation}
\tau_{\rm e, IL} \sim 3 \times 10^{7} n^{-1} {\rm ~yr}, 
\end{equation} 
where $n$ is the number density of ambient matter.
Hence, their life is expected to be 10$^{6}$ years at shortest 
in an interstellar medium.
The electrons will also lose their energy through inverse-Compton emission, 
and the life time of the electrons in a universe at $z$ for the inverse-Compton 
emission of the CMB photons is given as
\begin{equation}
\tau_{\rm e, IC} \sim 3 \times 10^{12} \gamma^{-1} (1+z)^{-4} {\rm ~yr}.  
\end{equation}
The lifetime of  electrons with $\gamma \sim 10^{3}$ for the inverse-Compton 
emission is longer than 10$^{6}$ years unless the electrons exist in a universe 
at $z$ close to 10.
Thus, we assume the lifetime of electrons responsible for the inverse-Compton 
emission in the X-ray band to be 10$^{6}$ years hereafter.

The X-ray spectrum of the inverse-Compton emission from a SNR with an age 
less than 10$^{6}$ years is given as 
(see equation 7.31 in \cite{rybicki79}) 
\begin{equation}
\frac{dL}{dE} = 
(8\pi^{2}r_{\rm e}^{2} / h^{3}c^{2}) C_{\rm e} F(\mu) (kT_{\rm CMB})^{(\mu+5)/2} 
E^{-(\mu-1)/2} ,
\end{equation}
where $r_{\rm e}$ and $h$ are the classical electron radius and 
the Planck constant respectively, and 
$F(\mu)$ is given as in Rybicki and Lightman (\yearcite{rybicki79}).
Thus, the X-ray luminosity of the inverse-Compton emission 
in an energy range between $E_{1}$ and $E_{2}$ is given by  
\begin{equation}
L(E_{1}, E_{2}) = 
(8\pi^{2}r_{\rm e}^{2} / h^{3}c^{2}) C_{{\rm e}} F(2) (kT_{\rm CMB})^{7/2} 
2 E_{1}^{1/2} [(E_{2}/E_{1})^{1/2} - 1)] ,
\end{equation}
for $\mu=2$.
If we assume $f=0.1, U_{\rm Th}=10^{51}$ erg, $T_{\rm CMB}=2.7$ K, $\mu=2$,  
and ln($\gamma_{\rm p, Max}$) = 10, 
the 1 -- 10 keV luminosity is estimated as 
\begin{equation}
L(1, 10) \simeq 6.2 \times 10^{30} {\rm ~erg~s^{-1}} 
\end{equation}
The inverse-Compton emission was really detected from SN 1006 
by a TeV $\gamma$-ray observation with the CANGAROO Cerenkov light 
telescope \citep{tanimori98}, 
although the Synchrotron emission is dominant over the 
inverse-Compton emission in the X-ray band \citep{koyama95}.
\citet{allen01} performed a joint spectral analysis of RXTE-PCA, ASCA-SIS, 
and ROSAT-PSPC data of SN 1006, and obtained $f U_{\rm Th} \sim 10^{50}$ erg s$^{-1}$, 
$\mu \sim 2$, and ln($\gamma_{\rm p, Max}) \sim 10$. 
These values are all consistent with those expected from general arguments 
on Galactic cosmic-ray acceleration (see discussion in \cite{allen01} 
and references therein).

\section{Emission from a Star-Bursting Proto-Galaxy} 

Let us consider how an X-ray flux of the inverse-Compton emission 
from a SNR evolves with the cosmic redshift factor, $z$.
Since $T_{\rm CMB, z} = T_{\rm CMB}(1+z)$ and 
$E_{\rm 1, z} = E_{\rm 1} (1+z)$, 
a flux of the inverse-Compton emission 
in an energy range between $E_{1}$ and $E_{2}$ detected 
from a SNR at $z$, 
$F(E_{1}, E_{2})_{\rm SNR, IC, z}$, is given as 
\begin{equation}
F(E_{1}, E_{2})_{\rm SNR, IC, z} = \frac{L(E_{1}, E_{2})(1+z)^{4}}
{4\pi r_{\rm L}^{2}},
\end{equation}
where $r_{\rm L}$ is the luminosity distance.

Here, we compare this $z$-dependence with that of the Synchrotron emission 
in the same X-ray band from the same electrons in the SNR.  
An X-ray spectrum of the Synchrotron emission is 
$\propto E_{\rm T}^{1/2} E^{-\mu/2}$ in an energy range, $E > E_{\rm T}$, 
where $E_{\rm T}$ is the turn-over energy corresponding to a turn-over 
Lorentz factor, $\gamma_{\rm T}$, above which the slope of 
the electron distribution becomes $\mu+1$.
$\gamma$ of electrons responsible for the Synchrotron emission in 
the X-ray band is as large as $\sim 10^{8}$, while $\gamma_{\rm T}$ 
in 10$^{6}$ years is calculated to be $\sim 3 \times 10^{6} (1+z)^{-4}$ 
from equation 5 on an assumption that the main energy loss of 
the high-energy electrons is due to  inverse-Compton emission.  
Since $E_{\rm T} \propto \gamma_{\rm T}^{2} kT_{\rm CMB} \propto (1+z)^{-7}$, 
the ratio of the flux of the inverse-Compton emission to that of the Synchrotron 
emission is proportional to $(1+z)^{15/2}$ for $\mu=2$.
Hence, the inverse-Compton emission should be dominant over the Synchrotron 
emission in the X-ray observations of relics of star-burst activities in 
distant galaxies.

The integrated flux from a proto-elliptical galaxy at $z$ is expected to be 
\begin{equation}
F(E_{1}, E_{2})_{\rm P.G., z} = F(E_{1}, E_{2})_{\rm SNR, z} 
N_{\rm SN} \tau_{\rm e} / \tau_{\rm SFB}. 
\end{equation}
Here, $\tau_{\rm e}$ is the average life of non-thermal electrons 
responsible for inverse-Compton X-ray emission, and is assumed to be 
10$^{6}$ years.  
$N_{\rm SN}$ and $\tau_{\rm SFB}$ are the total SN-number in a star-formation 
burst and the duration of the star-formation  burst.
They are roughly 10$^{10}$ and 
10$^{7.5}$ years, respectively, as in \citet{elbaz95}.

A result of calculations of equation (10) for $L(1, 10)=1\times10^{31}$ erg s$^{-1}$, 
$N_{\rm SN}=10^{10}$, $\tau_{\rm e}=10^{6}$ yr and 
$\tau_{\rm SFB}=10^{7.5}$ yr, is shown in figure 1 as a function of $z$.  
In these calculations, $\Omega = 0.3, 
\Lambda=0.7, H_{0}=$70 km s$^{-1}$Mpc$^{-1}$ are assumed.  It is very interesting to note 
that the observed flux is expected to rather increase as $z$ increases beyond 
unity.  This is simply due to an increase of the CMB flux in association 
with a $z$-increase.  
\begin{figure}
\begin{center}
\FigureFile(80mm,50mm){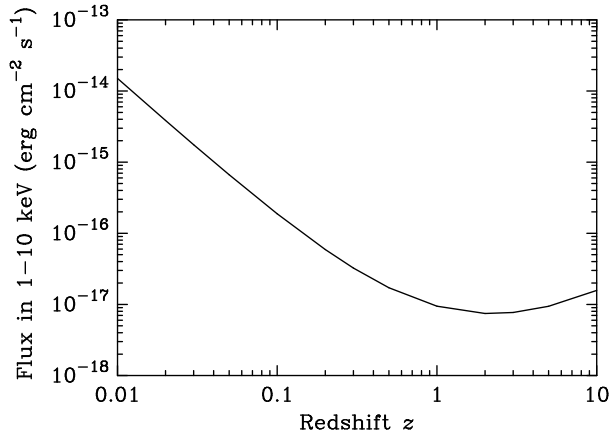}
\end{center}
\caption{Estimated X-ray flux from star-formation-bursting galaxies as a 
function of $z$.} 
\end{figure}

\section{Discussion}

As can be seen in figure 1, the expected flux in 1 -- 10 keV can exceed $10^{-17}$ 
erg cm$^{-2}$ s$^{-1}$ at $z \sim 10$.  If the parameters adopted above are 
really adequate, such an initial star-burst activity in proto elliptical 
galaxies could be detectable in a 10$^{5}$ s observation with a large X-ray 
telescope with a photon collecting area of several 10$^{5}$ cm$^{2}$, such as 
intended in the XEUS mission \citep{bavdaz99}. 
Such a large telescope could be realized in fairly near future.

According to results from optical deep survey observations 
(e.g., \cite{furusawa00}), the number of galaxies with $z > 1$ is 
about 10$^{4}$ deg$^{-2}$.  Even if such a distant galaxy has a violent 
star formation burst (SFB) with a period of 10$^{7.5}$ years as discussed above, 
a chance to hit the SFB in a galaxy will be only $\sim$ 1\%.  
Hence, if a fraction, of the order of 10\%, of the galaxies experience 
the violent SFBs, we could be able to detect the inverse-Compton X-rays 
from a number, of the order of 10, of star-bursting galaxies in a deg$^{2}$ field.

Such an X-ray source would have the  spectrum of a power law with a photon 
index of around 1.5. Although this spectrum is similar to that of AGNs, 
the star-burst activity in a proto galaxy could have a spatial extension of 
several ten kpc. If so, these sources should be detected as sources 
with an extension of several arc seconds, and hence be distinguishable from AGNs.

Even if we detect star-bursting proto galaxies, it might be difficult 
to determine their distances. One possibility to do it could be redshift 
determination of nuclear $\gamma$-ray lines from the SN ejectas.

In a type II SN, an Fe$^{56}$ atom is produced through a process, 
Ni$^{56}$ -- Co$^{56}$ -- Fe$^{56}$, and a Co$^{56}$ emits a 1.2 MeV 
or 0.85 MeV line in decaying with a time constant of 111 days. 
These nuclear lines are down-scattered by ambient cool electrons 
and are fully absorbed by the ambient matter at the initial phase 
of a SN explosion. However, as the ejecta expands, the ambient matter 
gradually becomes transparent for $\gamma$-ray lines in one or two years. 
At this stage, $\gamma$-ray lines from Co$^{57}$, 122 keV or 136 keV in a 
sequence, Ni$^{57}$--Co$^{57}$--Fe$^{57}$, are considered to be more 
detectable than those from Co$^{56}$. The amount of Fe$^{57}$ is only 
$\sim 5\% $ of that of Fe$^{56}$ but the time constant of the Co$^{57}$ 
decay to $Fe^{57}$ is as long as 391 days. Since this decay time is 
comparable to the time scale for the ejecta to become transparent, 
about 10\% of the 122 keV or 136 keV lines emerges without any interaction 
with the ambient matter (see e.g., \cite{kumagai93}).

The expected Co$^{57}$ line (a total of 122 keV lines and 136 keV lines) 
flux can be estimated as
\begin{equation}
\Phi_{{\rm Co-line}} \sim (f_{\rm det} f_{57/56} 
M_{\rm Fe}/m_{\rm Fe})(1+z)/\tau_{\rm SFB}
/(4\pi r_{\rm L}^{2}) ,
\end{equation}
where $f_{\rm det}, f_{57/56}, M_{\rm Fe}, m_{\rm Fe}$ are a probability for a 
Co$^{57}$ line to be directly detected, the abundance ratio of Fe$^{57}$ 
to Fe$^{56}$, the total iron mass produced in a star-formation burst, 
and the mass of an iron atom, respectively. The Co$^{57}$ line flux was 
calculated from equation (11) based  on assumptions of the following parameter values, 
$f_{\rm det} = 0.1$ (from \cite{kumagai93}); $f_{57/56}=0.04$ 
\citep{kurfess92}; $M_{\rm Fe}=10^{9} M_{\odot}$ (from \cite{arnaud92}),
as a function of $z$. The estimated photon flux of the Co$^{57}$ lines from 
 proto elliptical galaxies is about 2 -- 6 $\times 10^{-12}$ photons 
cm$^{-2}$ s$^{-1}$ for sources at $z \sim$ 5--10. This is 
far below the detection limit 
of such  large telescopes having a photon collecting area as large as 
10$^{4}$ cm$^{2}$ in 10 -- 20 keV (the Co$^{57}$ lines will appear 
in this range for sources with $z\sim$ 5--10). However, after picking up 
a sufficient number of serendipitous, slightly extended sources with 
AGN-like spectra, possible candidates of star-bursting proto elliptical 
galaxies, we might be able to sum up all of those spectra and to see the 
$z$-distribution of the Co$^{57}$ lines.

\end{document}